# Application of Adjoint-Based Optimal Control to Intergo-Differential Forward System

Kadyrova Alena, Khlyupin Aleksey
MIPT Center for engineering and technology
(presented at the ECMOR conference on 3.09.2018, Barcelona)

**Abstract**

In recent years researchers in oil-gas industry have established that the contribution of memory is significant for the modeling of fluid flow in unconventional reservoirs. According to modern works, the memory effect appears due to the contrast between high permeable fractures and nanoporous matrix leading to the gap of fluid velocities at the interface between these media. Also in the homogenization procedure from micro to macro scale the nonlocality in time (in other words, the memory) reflects the delay of fluid pressure and density between subdomains with different properties of pore space geometry.

Mathematically, a memory-based fluid flow model can be described by the system of integro-differential equations. Despite the fact that a large number of journal articles are devoted to numerical methods for the forward solution of such equations, the problems of optimization and optimal control of these systems are actual and insufficiently studied.

We consider the one-dimensional model of gas filtration and diffusion as a model with memory. The system includes a partial differential equation for filtration in fractures and weakly singular Volterra integral equation of the second kind, which describes the diffusion of gas from blocks with closed nanopores. Numerical simulation, obtained using a Navot-trapezoidal algorithm, shows that the effect of memory influences on the distribution and the time evolution of pressure and density in comparison with the classical double porosity model.

The pressure-constrained maximization of discounted cumulative gas production was chosen as a basic optimization problem. The appearance of memory in the model makes the standard adjoint-based approach not applicable since it was developed only for conventional systems of partial differential equations. The novel adjoint model for media with memory was obtained from the necessary conditions of optimality using the classical theory of calculus of variations and efficiently applied to production optimization problem.

In conclusion we compare optimal control scenarios for the model with memory and for the classical double porosity model. Analysis has shown the importance of memory accounting in reservoir optimization problems. Also the results of parametric analysis are considered and discussed.

**Introduction**

A wide variety of reservoir management issues in the oil and gas industry can be addressed through the application of modern optimization and optimal control theory. Thus, the practical importance and necessity of research in this area of applied mathematics is difficult to overestimate. In planning new wells, key questions include the determination of the optimum well location, well type and drilling schedule, given operational and economic constraints. For existing wells, significant benefit can be achieved through optimization of well settings such as bottom hole pressures (BHPs) or flow rates, as a function of time.

Practical production optimization problems typically involve large, highly complex reservoir models, thousands of unknowns and many nonlinear constraints, which makes the numerical calculation of gradients for the optimization process impractical. In the one of the first works on this topic (Sarma et al. 2005) authors explore the adjoint algorithm for production optimization for the conventional Darcy flow problem using optimal control theory. The main approach that is still alive is to use the underlying simulator as the forward model and its adjoint for calculation of gradients. In the review paper (Jansen and Durlofsky 2017) authors focus on well control, which entails the determination of optimal well settings that maximize a particular objective function. As is the case with most reservoir-related optimizations, this problem is in general computationally demanding since function evaluations require reservoir simulation runs. Here they describe reduced-order modeling procedures, which act to accelerate these simulation runs, and discuss their use within the context of well control optimization.

Also, many other aspects of reservoir management can be expected to benefit from the application of computational optimization procedures. The key ingredients to successful real-time reservoir management, also known as a «closed-loop» approach, include efficient optimization and model-updating algorithms, as well as techniques for efficient uncertainty propagation. In data assimilation, one combines the information present in the observations and an existing numerical model, as well as the uncertainties in the observations and the models, to produce more realistic results. Data assimilation is required in several fields including meteorology (Talagrand et al. 1987), oceanography, groundwater flow (Valstar et al. 2004) and petroleum engineering. In the latter, it is known as history matching.

Gradient-based history matching algorithms can be used to adapt the uncertain parameters in a reservoir model using production data. In the work (Kaleta et al. 2011) a novel approach to gradient-based history matching which is based on model reduction was proposed. The original (nonlinear and high-order) forward model is replaced by a linear reduced-order forward model. The reduced order model is constructed with the aid of the proper orthogonal decomposition method. Authors of the work (Sarma P. et al. 2006) discussed a simplified implementation of the closed-loop approach that combines efficient optimal control and model-updating algorithms for real-time production optimization.

As follows from the above, a large amount of work has been done for the classical filtration problems which are described by partial differential equations. However, in recent years researchers in oil-gas industry have established that the contribution of memory is significant for the modeling of fluid flow in unconventional reservoirs. Mathematically, the diffusion equation with memory has a form of Volterra integral equation of the second kind with a singular kernel and is strongly nonlinear. The previously developed adjoint algorithms can not be directly applied to equations of this type. Thus the development of a new gradient-based adjoint technique is highly desirable.

Let's consider the characteristic features of unconventional reservoirs, which require the development of a new approach for flow modeling. The main features of reservoirs of this type are ultra-low permeability and high heterogeneity caused by different scales of connected and isolated fractures and pores. The geological heterogeneity of this system leads to different flow mechanisms appearing at various time and space scales. In the case of a gas flow in unconventional reservoirs, Javadpour et al. (Javadpour et al. 2007) identified five flow mechanisms which contribute to total production, ranging from a molecular scale (slow local mechanism) to a macroscale (fast global mechanism).

According to modern works (Panfilov and Rasoulzadeh 2013), (Panfilov and Fourar 2006), the memory effect appears due to the contrast between high permeable fractures and nanoporous matrix leading to the gap of fluid velocities at the interface between these media. Also in the homogenization procedure from micro to macro scale the nonlocality in time (in other words, the memory) reflects the delay of fluid pressure and density between subdomains with different properties of pore space geometry.

Thus, it seems questionable that the models developed for homogeneous continuous media are able to describe fluid motion in such type of reservoirs correctly. Instead, such systems can be better represented as random, heterogeneous environments in which diffusion does not conform to traditional transfer laws.

In a paper (Miah et al. 2017) the authors note that the memory effect can be an efficient tool for accurate description of the fluid and rock properties that vary with time and space. The work of Holy (2016) is devoted to the development of numerical methods for solving the anomalous diffusion equation for linear regime of flow in Stimulated Reservoir Volume (SRV). The expansion of the model for multiphase flow was proposed, but the model of anomalous diffusion for a multiphase flow requires an additional research activity. In addition, in this paper authors propose a technique that makes it possible to apply anomalous diffusion for analyzing field data and predicting well production rates in unconventional reservoirs.

In the work (Hossain 2016) the finite-difference approximation of the diffusion equation with memory was developed. A model of diffusion with memory was numerically realized, a comparison was made with the traditional model of filtration using the example of actual field data obtained from the field in the Middle East. The results obtained by the authors demonstrated that the model with a memory describes real data much more better than the model based on the Darcy flow.

Even though a several number of journal articles are devoted to numerical methods for the forward solution of such equations, the problems of optimization and optimal control of these systems are actual and insufficiently studied. This is the case when the physical essence of the model significantly affects the structure of the adjoint equations due to the appearance of an integral term in the governing equations reflecting the memory effect. Thus, the development of optimization methods for systems with such physics is an actual challenge and will contribute to the next step towards practical application of applied mathematics to smart reservoir management.

In this paper we have considered a simple model of diffusion with the memory effect. We performed a numerical simulation of the model and compared pressure control scenarios for the model with memory and for the classical double porosity model. The results of the numerical realization showed that there are differences in the pressure distributions obtained from both cases. Thus, it was shown the importance of memory accounting in reservoir optimization problems. The pressure-constrained maximization problem of gas production was considered. The appearance of memory in the model makes the standard adjoint-based approach not applicable since it was developed only for conventional systems of partial differential equations. Finally we provide a complete derivation of novel adjoint method for the efficient computation of the gradients for models with such physics and verify it on several cases.

This paper is arranged as follows: forward model equations are briefly discussed in the next section. Then we formulate new version of adjoint model for chosen optimization problem. The results are presented in the last section. Through several examples we demonstrate that the proposed novel approach provides a practical strategy for optimal control that depends on rock and fluid properties which describe the relative contribution of memory to the cumulative gas flow.

**Model formulation**

In this section we investigate the influence of memory effect at the gas flow in the heterogeneous media. We choose the one-dimensional single-phase model of gas filtration and diffusion proposed in paper (Alekseev et al. 2007) as the example of integro-differential models. This model represents simplified flow and satisfies all desired properties, that makes it suitable for our analysis.

In this model the reservoir is considered as the layer of spherical blocks separated by the system of connected fractures and open pores; disconnected closed pores inside the blocks represent the main storage volume of desired gas we want to produce. At equilibrium, gas molecules occupy pores as compressed gas and disperse in the matrix blocks as dissolved gas. The relation between the gas concentration in a solid phase and the density of the gas phase is established by the Henry's law. Gas start flowing toward the low pressure zone since the equilibrium is disturbed during borehole drilling or by fracture inducing.

The gas flow in considering system is governed by two physical processes: the filtration in fractures and the diffusion from blocks. Production starts from the open fracture network, then pressure drop in the fractures leads to the diffusion of gas molecules within the matrix blocks towards the pore walls. The main flow occurs in the fractures, while blocks play role of gas sources for them. The characteristic times of flow processes are significantly different for fractures and for blocks.

The system includes a partial differential equation for filtration in fractures and weakly singular Volterra integral equation of the second kind for the diffusion of gas from blocks with closed nanopores (1). The filtration is governed by Darcy's law. The diffusion equation reflects the memory effect in this media.

$$\begin{cases} \dfrac{\partial}{\partial t}\left[\gamma_0 \rho(x,t) + (1-\gamma_0)\left(1-\gamma+\dfrac{\gamma}{v}\right)c(x,t)\right] = kT\dfrac{\partial}{\partial x}\left[\dfrac{\rho(x,t)}{\mu}\dfrac{\partial \rho(x,t)}{\partial x}\right] \\ c(x,t) = c_0 - \dfrac{3}{R}\sqrt{\dfrac{D}{\pi\left(1-\gamma+\dfrac{\gamma}{v}\right)}}\int_0^t \dfrac{c(x,\tau)-v\rho(\tau)}{\sqrt{t-\tau}}d\tau \end{cases} \quad (1)$$

where $\rho(x,t)$ – gas density in fractures, $c(x,t)$ – concentration in blocks, $c_0$ – initial concentration, $\gamma_0$ – open porosity, $\gamma_0$ – closed porosity, $k$ – permeability of fractures, $v$ - solubility, $T$ - temperature, $\mu$ – viscosity, $R$ – radius of blocks, $D$ – diffusivity coefficient.

**Numerical solution**

For the numerical simulation of considering system the fully implicit finite-difference scheme is developed and efficiently applied. Modeling of nonlocal, memory-dependent flow requires the application of special techniques for solving the integral equation with weakly singularity. For this purpose we use Navot's quadrature rule, which is applied for functions having a singularity of any type on or near the endpoints of the integration interval. This algorithm provides enough high accuracy and satisfactory approximation. In addition, the computational complexity of this method is relatively low since it doesn't require additional work to compute integration weight coefficients. This fact is significant for the simulation of large dimensional hydrodynamic models.

Navot's quadrature are applied for functions having a singularity of any type on or near the endpoints of the integration interval, which can be written in following general form:

$$I(G) = \int_a^b G(x)dx = \int_a^b (b-x)^\alpha g(x)dx \quad (2)$$

where $-1 < \alpha < 0$ and $G(x) = (b-x)^\alpha g(x)$, $g(x)$ is a smooth function on the interval $[a,b]$.
This method assigns equal weights to the endpoints and arbitrary weights to the points in the integration interval with equal distance between the adjacent points. The integrand at a singular point is expressed in terms of the Riemann-Zeta function using the Euler Maclaurin summation formula (see reference (Navot 1961) for detailed explanation).

For the approximation of the singular integral in diffusion equation we apply the modified trapezoidal rule $T_h(G)$, which has been introduced in (Liu and Tao 2007) by using Navot's quadrature as follows:

$$T_h(G) = \dfrac{h}{2}G(x_0) + h\sum_{j=1}^{N-1} G(x_j) - \zeta(-\alpha)g(b)h^{1+\alpha} \quad (3)$$

where step length $h = \dfrac{b-a}{N}$ and mesh points $x_j = a + jh$, $j = 0,\dots N$, where $N$ is a sufficiently large number, $\zeta(-\alpha)$ is the Riemann-Zeta function. The error estimate in Navot–trapezoidal rule is $O(h^{3+\alpha})$.

**Results**

In this section we discuss the comparative analysis of the model with memory and the classical double porosity model, which is widely used for the description of fractured porous media.

Classical double porosity model is developed first in (Barenblatt et al. 1960). It is assumed that the medium consists of two overlapping continua: highly permeable fractures and low-permeable matrix. The interaction between them is modeled through a semi-empirical transfer term.

*Table 1 Reservoir and fluid properties, operational parameters*

| Temperature, $T$ | 360K |
|---|---|
| Solubility, $v$ | $10^{-1}$ |
| Radius of blocks, $R$ | $10^{-3}$ m |
| Viscosity, $\mu$ | $10^{-5}$ Pa·s |
| Closed porosity, $\gamma$ | 0.5 |
| Open porosity, $\gamma_o$ | 0.1 |
| Permeability of fractures, $k_f$ | 1 D |
| Geometric coefficient of porous media, $\alpha$ | $10^{-11}$ |
| Pressure at the right boundary, $p_0$ | 30 MPa |
| Bottom hole pressure, $p_{BHP}$ | 20 MPa |
| Perturbed pressure, $p_1$ | 30 MPa |

In order to investigate the impact of memory the following case is simulated. We consider the one-dimensional reservoir of size 10 m with constant pressure right boundary $p_0$. We take the uniform grid mesh of 10 blocks. There is a production well on the left boundary, which operates at constant bottom hole pressure $p_{BHP}$ during time *t*. Before the occurrence of steady state flow the BHP over a time *dt* the BHP is perturbed and then the model is evaluated again. The schematic of BHP perturbation is shown in Figure 1. We assume, that the external perturbation directly affects only on the pressure in fractures, which instantly reacts on this perturbation. The reservoir and fluid properties used in simulation are given in Table 1.

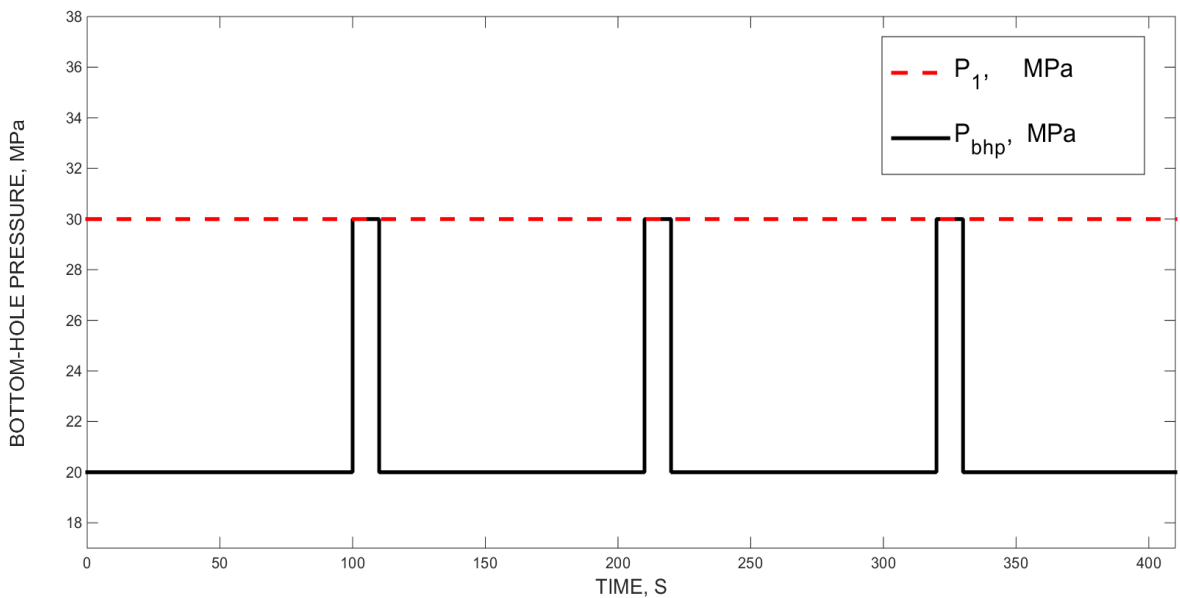

*Figure 1 Schematic of BHP perturbation*

The numerical simulation of this problem is carried out for the classical double porosity model and for the memory-based model with different values of the diffusivity coefficient. The results is presented in the form of the pressure-time plots for fixed grid cell (Figure 2).

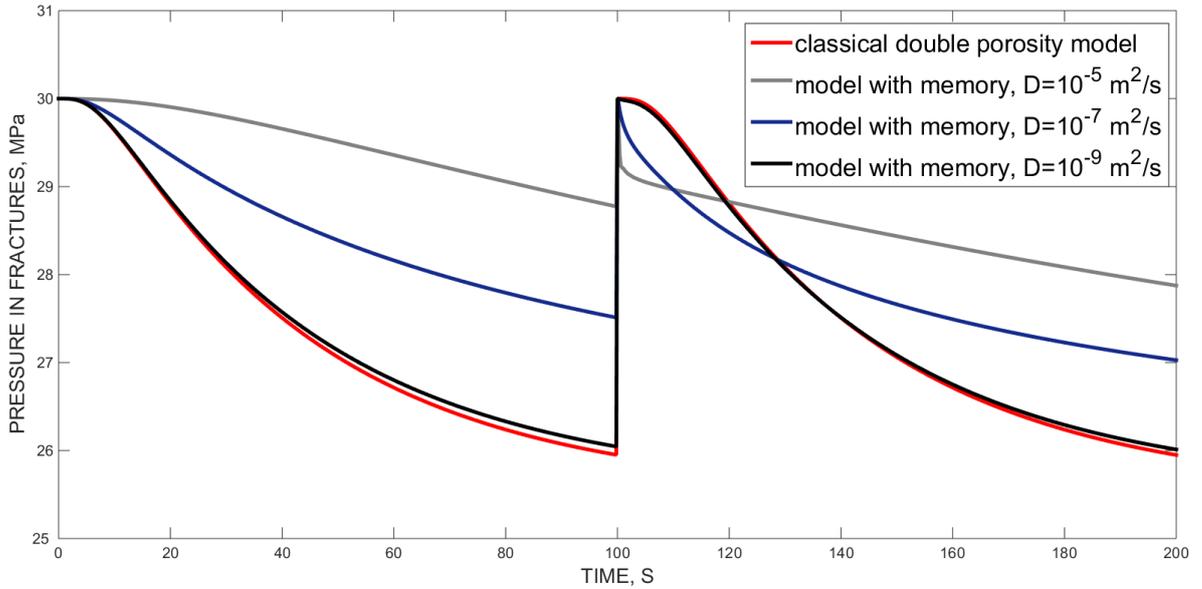

*Figure 2 Pressure in fractures as function of time*

- The analysis of obtained results clearly demonstrates the impact of memory and corresponding parameters on the distribution and the time evolution of pressure and density. We can draw followings conclusion based on the interpretation of the pressure-time plots.

- The classical double porosity model doesn't include the memory effect, so the pressure drop after external perturbation is similar to the initial pressure drop. The external perturbation can be characterized as the setting of new initial conditions.

- The pressure drop of the memory-based model after the external perturbation is different from the initial one. The history of the processes influences the subsequent development of the system.

- The nature of the pressure drop is influenced by the value the diffusivity coefficient. With the increase in the diffusivity coefficient, the rate of pressure drop in the fractures decreases, which can be explained by the fact that fractures are faster supplied by gas molecules diffusing from the blocks.

- With the decrease of the diffusivity coefficient, the pressure function of memory-based model converges to the pressure function of the classical double porosity model.

These statements prove the importance of memory accounting for the modeling of gas flow. The various characteristics of obtained pressure drop curves make suppose that optimal controls for models with and without this new physical aspect can be significantly different. Changes in well settings leads to the difference in cumulative production estimation, so the production optimization in such media is of interest.

**Optimization problem formulation**

In this work we consider the basic optimization problem, which can be described with following mathematical formulation (4). The objective is to maximize the cumulative gas production $J$ in the fixed forecast period. The cost function is the summation over all time steps the kernel $L^n$, which includes the average gas production $q_g^{n+1}$ for $\Delta t^n$ time interval and the expression in the denominators, that represents the discounting effect, where r is the discount rate in percent. The gas flow in the reservoir is determined by initial conditions together with fully implicit forward model $g^n(x^{n+1}, x^n, .., x^1, x^0, u^n)$, which is presented by the system of integro-differential equations for each time step (1). Here $x^n$ refers the dynamic states of the system, in our case $x^n$ contains the concentration in blocks and the density in fractures.

The production well is set on the left boundary of the one-dimensional reservoir, the bottom hole pressure could be manipulated in order to achieve the optimum, so the bottom hole pressure $p_{bhp}^n$ is considered as control parameter $u^n$. The boundary condition at the right hand side corresponds to no-flow regime.

Maximization of the cost function is carried out under additional constraints for the controls. We assume that the only production constraints are upper ($UB$) and lower ($LB$) bounds on the bottom hole pressure $p_{bhp}^n$.

$$\max_{u^n}\left[J(u) = \sum_{n=0}^{N-1} L^n(x^{n+1}, u^n) = \sum_{n=0}^{N-1} q_g^{n+1} \Delta t^n\right]$$
$$\text{subject to } g^n(x^{n+1}, x^n, .., x^1, x^0, u^n) = 0, \quad \forall n \in (0, ..N-1)$$
$$x^0 = x_0, \quad (Initial\ Condition)$$
$$LB \leq u^n \leq UB, \quad \forall n \in (0, ..N-1)$$
(4)

Thus, the problem described above requires finding a sequence of bottom hole pressure $p_{bhp}^n$ for $n = 0, ..N - 1$, where $n$ is the time step index and $N$ is the total number of time steps, to maximize the cost function $J(u^0, ..u^{N-1})$.

The most commonly used methods for solving production optimization problems are iterative gradient algorithms. The application of gradient-based procedures to practical reservoir management requires an efficient technique to calculate the total derivatives of a cost function with the respect to control parameters, since depending on the chosen algorithm it may be computationally very expensive due to complexity and the large dimension of hydrodynamic models. The adjoint methods for gradient calculation are proved to be the most efficient among existing approaches.

Adjoint-based algorithms are successfully applied to optimal control of real reservoirs with large simulation models and different types of constrains (Sarma et al. 2005), (Brower and Jansen 2002). The application area of adjoint method is not limited by production optimization, it is also used in other important problems of oil-gas industry, such as history matching (Sarma et al. 2006), (Oliver et al. 2008), (Zakirov et al. 2015).

According to classical works on the optimal control of production development, gradient is obtained by solving the adjoint system. The standard adjoint model can be written in following form:

$$\begin{cases} \lambda^{Tn} = -\left[\frac{\partial L^{n-1}}{\partial x^n} + \lambda^{T(n+1)}\frac{\partial g^n}{\partial x^n}\right]\left[\frac{\partial g^{n-1}}{\partial x^n}\right]^{-1}, & n \in (0, ..N-1) \\ \lambda^{TN} = -\frac{\partial L^{N-1}}{\partial x^N}\left[\frac{\partial g^{N-1}}{\partial x^N}\right]^{-1}, & (Final\ Condition) \end{cases}$$
(5)

The vector λ is often referred to as the adjoint, co-state or Lagrange multiplier.

This standard adjoint system is developed only for conventional systems of partial differential equations, which suppose that equations for any time step depend on two consecutive dynamic states.

The main characteristic of considering forward model is the appearance of memory. It means that on each time step n the system of equations $g^n(x^{n+1}, x^n, ..x^0, u^n)$ depends on all previous dynamic states. The integro-differential forward system makes the standard adjoint-based approach not applicable since it was obtained for traditional hydrodynamic models.

**Adjoint model for memory-based system**

We derive the novel adjoint model for media with memory using the Lagrange formalism. Basing on the classical approach proposed in paper (Sarma et al. 2005) the adjoint model for media with memory is obtained from the necessary conditions of optimality using the classical theory of calculus of variations.

We consider the general form of the cost function with only simulation constraints and introduce the Lagrangian corresponding to our optimization problem (6).

$$J_A = \sum_{n=0}^{N-1} L^n(x^{n+1}, u^n) + \lambda^{T0}(x_0 - x^0) + \sum_{n=0}^{N-1} \lambda^{T(n+1)} g^n(x^{n+1}, x^n, \ldots x^0, u^n) \qquad (6)$$

The first variation of $J_A$ is given by:

$$\begin{aligned}
\delta J_A &= \sum_{n=0}^{N-1} \frac{\partial L^n}{\partial x^{n+1}} \delta x^{n+1} + \sum_{n=0}^{N-1} \frac{\partial L^n}{\partial u^n} \delta u^n + (x_0 - x^0)\delta\lambda^{T0} + \sum_{n=0}^{N-1} [g^n]\delta\lambda^{T(n+1)} + \\
&+ \sum_{n=0}^{N-1} \left[\sum_{k=n}^{N-1} \lambda^{T(k+1)} \frac{\partial g^k}{\partial x^{n+1}}\right] \delta x^{n+1} + \sum_{n=0}^{N-1} \lambda^{T(n+1)} \frac{\partial g^n}{\partial u^n} \delta u^n = \\
&= \sum_{n=1}^{N} \left[\frac{\partial L^{n-1}}{\partial x^n} + \sum_{k=n}^{N} \lambda^{Tk} \frac{\partial g^{k-1}}{\partial x^n}\right] \delta x^n + \sum_{n=0}^{N-1} \left[\frac{\partial L^n}{\partial u^n} + \lambda^{T(n+1)} \frac{\partial g^n}{\partial u^n}\right] \delta u^n
\end{aligned} \qquad (7)$$

For optimality the first variation of Lagrangian must equal zero. The total variation is observed as the sum of variations of $x^n$, $u^n$ and $\lambda^{Tn}$. Terms $[g^n]\delta\lambda^{T(n+1)}$ and $(x_0 - x^0)\delta\lambda^{T0}$ are zero by definition. A further simplification of (7) can be obtained through putting a restriction on the Lagrange multipliers $\lambda^{Tn}$ as follows:

$$\begin{cases} \lambda^{Tn} = -\left[\frac{\partial L^{n-1}}{\partial x^n} + \sum_{k=n+1}^{N} \lambda^{Tk} \frac{\partial g^{k-1}}{\partial x^n}\right]\left[\frac{\partial g^{n-1}}{\partial x^n}\right]^{-1}, \quad n \in (0, \ldots N-1) \\ \lambda^{TN} = -\frac{\partial L^{N-1}}{\partial x^N}\left[\frac{\partial g^{N-1}}{\partial x^N}\right]^{-1}, \qquad (Final\ Condition) \end{cases} \qquad (8)$$

The system (8) represents the adjoint model for media with memory. Note, that Lagrange multiplier $\lambda^{Tn}$ depends on $\lambda^{Tn+1}, \ldots, \lambda^{TN}$ and the costate vector for the last control step must be calculated first. Thus the novel adjoint system as well as the forward system has the memory effect, but it must be solved backwards in time.

With Lagrange multipliers calculated by (8), the variation of $J_A$ (9) and the required gradient of the cost function with the respect to the control parameters (10) have the same form as in the case of the standard adjoint method:

$$\delta J_A = \sum_{n=0}^{N-1} \left[\frac{\partial L^n}{\partial u^n} + \lambda^{T(n+1)} \frac{\partial g^n}{\partial u^n}\right] \delta u^n \qquad (9)$$

$$\frac{dJ}{du^n} = \frac{dJ_A}{du^n} = \left[\frac{\partial L^n}{\partial u^n} + \lambda^{T(n+1)} \frac{\partial g^n}{\partial u^n}\right], \qquad \forall n \in (0, \ldots N-1) \qquad (10)$$

**Examples**

We present the numerical examples demonstrating the implementation of the novel adjoint model with memory to the production optimization problem (4).

In order to illustrate the benefit of the optimization process, it is usual practice to compare the results of optimal control against the reference case. We choose a scenario in which the well is operated on constant bottom hole pressure as reference case.

Reservoir length is 10 m, modeled with 10 grid cells. Total production time is fixed per considering problem; it is set equal to 100 s and 50 time steps are used. Production starts after 5 time steps. The reservoir and fluid properties using in the example are the same as those given in Table 1. The diffusivity coefficient are taken to be $10^{-5}$ m$^2$/s. We set the upper bound of BHP equal to 30 MPa, the lower bound – 20MPa, the bottom hole pressure in the reference case – 23MPa, the initial reservoir pressure – 30 MPa.

The reference scenario serves as initial control for the optimization. The optimal solution of problem (4) is found by iteratively improving upon the initial choice of control parameters in a steepest descend method. Figure 3 shows results of this optimal control problem.

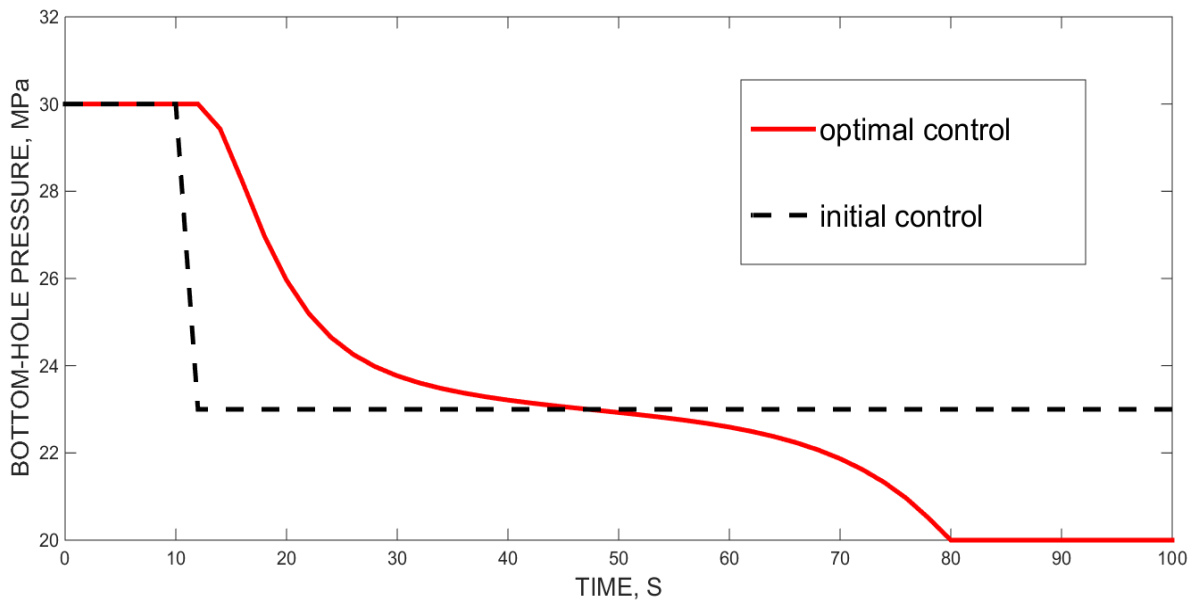

*Figure 3 Bottom hole pressure as function of time for optimized and base case*

In Figure 4 we compare the cumulative gas production of the optimized case and the base case. Note, that the value of cumulative production at the end of considering period for the optimal control is higher than this one for the initial control. The increase in the cumulative gas production (the objective function) in iterative process is shown in Figure 5. Each iteration corresponds to a backward adjoint run, producing gradient, followed by one or more forward simulation runs in order to the calculating of objective function. It is clear that the objective function increases in each iteration and converges to the optimized value with improvement over the reference case value. Therefore, we can conclude, that the local optimum is found.

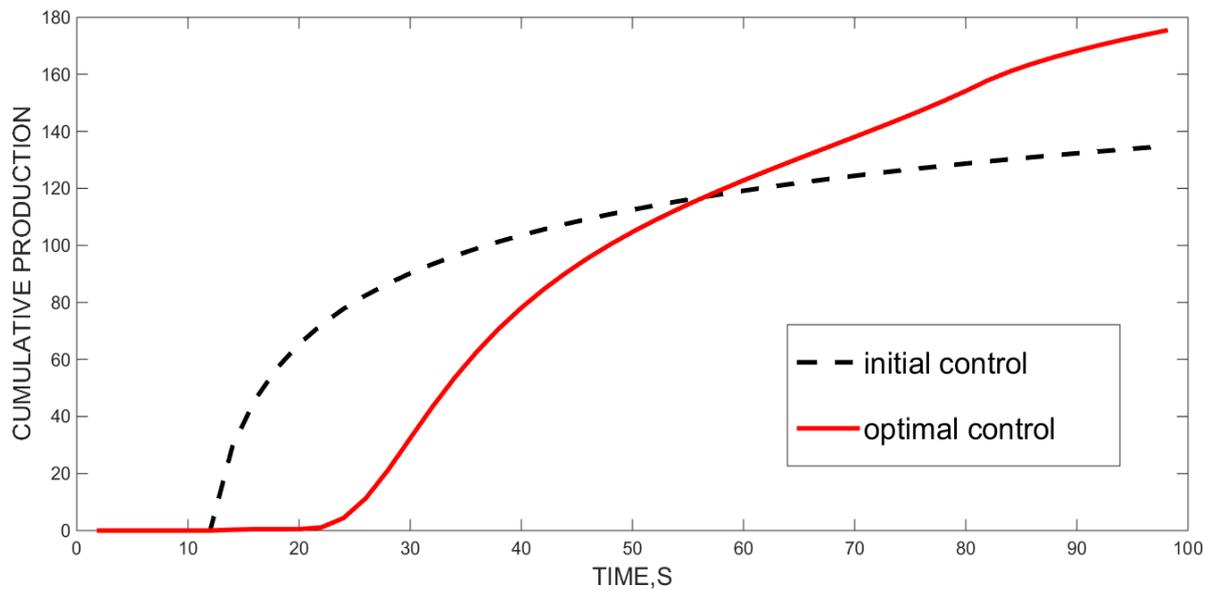

*Figure 4* *Cumulative production comparison of optimized and base case*

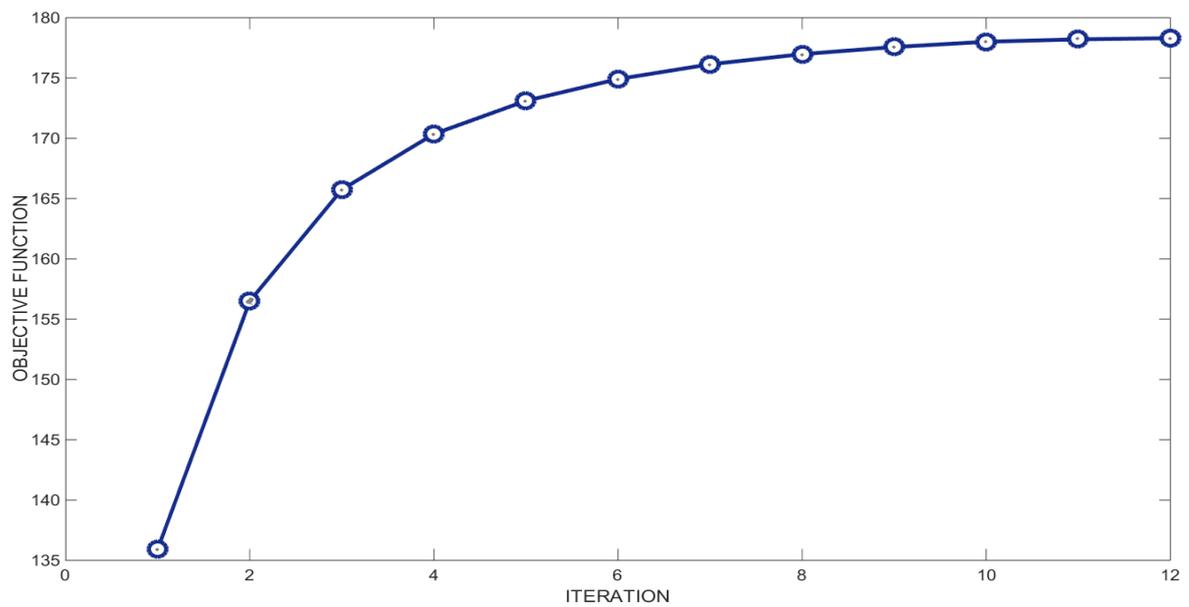

*Figure 5* *Convergence to optimum*

Since gradient methods in general provides ending up in local optimum, obtained solution should be considered as a minimal improvement and better control may be found.

In order to investigate the impact of model's parameters on optimal control we solve this optimization problem for the different values of the diffusivity coefficient ranged from $10^{-5}$ m$^2$/s to $10^{-9}$ m$^2$/s. The value of diffusivity coefficient reflects the weight of the relative contribution of memory effect to gas flow in forward model (1). Remind, that when the diffusivity coefficient is equal to $10^{-9}$ m$^2$/s considering nonlocal model converges to classical double porosity model without memory effect. Figure 6 demonstrates the optimal control strategies for the highest ant the lowest used values of diffusivity coefficient. Starting from the same initial controls and using the same reservoir and fluid properties, the optimization process for different parameters of the forward model results in different optimized BHP's trajectories. Therefore, it shows the importance of memory accounting for the solving inverse problems.

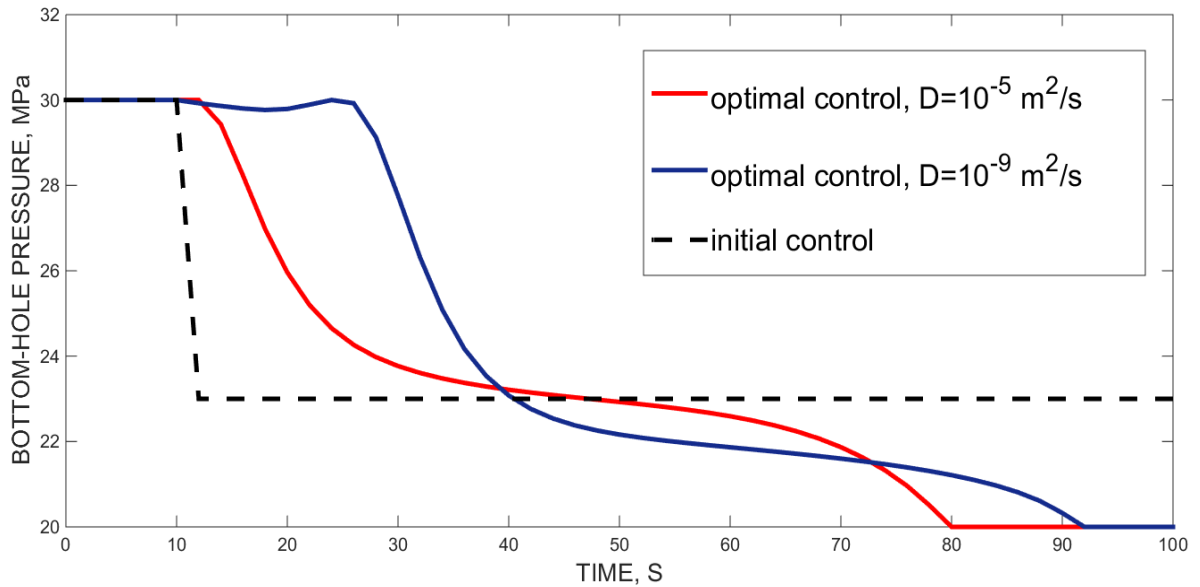

*Figure 6 Bottom hole pressure as function of time for different values of diffusivity coefficient*

We should note that the production for the smaller value of diffusivity coefficient starts later. It is caused by following facts. Figure 1 demonstrates the higher rate of pressure drop in the fractures for smaller diffusivity coefficient, and using ratio between the characteristic size of reservoir model and the simulation time results in pressure depletion. So the production of all fluid volume in the model with smaller diffusivity coefficient is faster, and choosing objective function is insensitive to the discounting effect.

**Conclusions**

In this paper we investigated the integro-differential model of gas filtration and diffusion. Numerical simulations have confirmed the significant contribution of memory to fluid flow in unconventional gas reservoirs.

The adjoint system for a model with memory effect has been obtained and efficiently implemented to the optimal control problem. The gradient-based optimization with using the proposed adjoint model resulted in an increase in the cumulative gas production with respect to the reference case.

Through the example we demonstrated the importance of memory accounting for the production optimization. The novel adjoint system with memory provided an optimal strategy that depends on model's parameters, which represent the relative contribution of memory effect to gas flow.

We should underline that although in this paper we considered the one-dimensional simplified gas flow, our novel adjoint system has been obtained in general form and could be implemented to production optimization and history matching for any memory-based forward model.

In this study we focused on the obtaining of qualitative results and considered the small dimension hydrodynamic model, but using of fully implicit finite-difference scheme and Navot's quadrature for approximating of weakly singular integral give an opportunity to increase the model characteristic size and time step with high accuracy.